# CAB: Towards the RNA-world in the interstellar medium – detection of urea, and search of 2-amino-oxazole and simple sugars

Izaskun Jiménez-Serra[1*], Jesús Martín-Pintado[1], Víctor M. Rivilla[2], Lucas F. Rodríguez Almeida[1], Elena R. Alonso Alonso[3,4], Shaoshan Zeng[5], Emilio J. Cocinero[3,4], Sergio Martín[6,7], Miguel Requena-Torres[8], Rafa Martín-Domenech[9], and Leonardo Testi[10,2]

[1]Departamento de Astrofísica, Centro de Astrobiología (INTA-CSIC), Carretera de Ajalvir km 4, Torrejón de Ardoz, 28850, Madrid, Spain; [2]INAF-Osservatorio Astrofisico di Arcetri, Largo Enrico Fermi 5, I-50125 Florence, Italy; [3]Departamento de Química Física, Facultad de Ciencia y Tecnología, Universidad del País Vasco, (UPV-EHU), 48080 Bilbao, Spain; [4]Biofisika Institute (CSIC, UPV/EHU), 48940 Leioa, Spain; [5]School of Physics and Astronomy, Queen Mary University of London, E1 4NS London, United Kingdom; [6]European Southern Observatory, Alonso de Córdova, 3107, Vitacura, Santiago, 763-0355, Chile; [7]Joint ALMA Observatory, Alonso de Córdova, 3107, Vitacura, Santiago, 763-0355, Chile; [8]Department of Astronomy, University of Maryland, College Park, MD, 20742, USA; [9]Harvard-Smithsonian Center for Astrophysics, 60 Garden Street, Cambridge, MA 02138, USA; [10]European Southern Observatory, Karl-Schwarzschild-Str. 2, D-85748 Garching bei München, Germany.

*To whom correspondence should be addressed: Izaskun Jiménez-Serra

ijimenez@cab.inta-csic.es. Tel: +34-915201071, Fax: +34-915201074

**Abstract**

In the past decade, Astrochemistry has witnessed an impressive increase in the number of detections of complex organic molecules. Some of these species are of prebiotic interest such as glycolaldehyde, the simplest sugar, or amino acetonitrile, a possible precursor of glycine. Recently, we have reported the detection of two new nitrogen-bearing complex organics, glycolonitrile and Z-cyanomethanimine, known to be intermediate species in the formation process of ribonucleotides within theories of a primordial ribonucleic acid (RNA)-world for the origin of life. In this paper, we present deep and high-sensitivity observations toward two of the most chemically rich sources in the Galaxy: a Giant Molecular Cloud in the center of the Milky Way **(G+0.693-0.027)** and a proto-Sun **(IRAS16293-2422 B). Our aim is** to explore whether the key precursors considered to drive the primordial RNA-world chemistry, are also found in space. **Our high-sensitivity observations reveal that urea is present in G+0.693-0.027 with an abundance of ~5x10$^{-11}$. This is the first detection of this prebiotic species outside a star-forming region. Urea remains undetected toward the proto-Sun IRAS16293-2422 B (upper limit to its abundance of ≤2x10$^{-11}$).** Other precursors of the RNA-world chemical scheme such as glycolaldehyde or cyanamide are abundant in space, but key prebiotic species such as 2-amino-oxazole, glyceraldehyde or dihydroxyacetone are not detected in either source. Future more sensitive observations targeting the brightest transitions of these species will be needed to disentangle whether these large prebiotic organics are certainly present in space.

**1. Introduction**

The question of the origin of life has intrigued human beings for centuries. Life appeared on Earth about 4 billion years ago, but we do not know the processes that made it possible. One of the proposed scenarios is the so-called ribonucleic acid RNA-world, which suggests that early forms of life relied solely on (RNA) to store genetic information and to catalyze chemical reactions. The RNA-world hypothesis was originally not very popular among the Astrobiology community because it was far from trivial to form RNA from its main constituents: a nitrogenous base, a ribose sugar and a phosphate. Even the prebiotic formation of these basic building blocks of the RNA is still, so far, unknown. In recent years, however, it has become clear that ribonucleotides could have formed, instead, from simpler molecules such as cyanamide ($NH_2CN$), cyanoacetylene ($HC_3N$) or glycolaldehyde ($CH_2(OH)CHO$), all derivatives of hydrogen cyanide (HCN) in an aqueous medium (Powner *et al.* 2009; Patel *et al.* 2015).

Astrochemistry is the discipline that studies the chemical processes taking place in space. Since the detection of the first molecular ions and radicals in space in the 1930s and 1940s (as e.g. CN, CH and $CH^+$; Swings & Rosenfeld 1937; Douglas & Herzberg 1941), Astrochemistry has witnessed an impressive increase in the number of molecular species detected in the gas phase in both the circumstellar medium (CSM, in particular around evolved stars), and in the interstellar medium (ISM, especially in regions forming stars). The average detection rate of new molecules in the CSM/ISM is currently about 4 species per year, and over 200 gas-phase molecular species are known to be present in space (see Figure 1 in McGuire 2018).

Among the detected molecules, Complex Organic Molecules (or COMs, defined as carbon-bearing compounds with 6 atoms or more in their molecular structure; Herbst & van Dishoeck 2009) have attracted much interest in recent years due to their link to prebiotic chemistry. Molecular searches in the ISM have indeed been very prolific in the detection of molecules of prebiotic relevance with relatively high abundances such as **cyanamide (Turner *et al.* 1975), glycolaldehyde (Hollis *et al.* 2000) or even urea (NH₂CONH₂; Belloche *et al.* 2019),** which hints to a high level of chemical complexity in space. The question that follows is: Should one expect to find in the ISM **the same chemical precursors as those of the** RNA-world hypothesis for the origin of life? Recently, we have reported the first detection of glycolonitrile in space (or hydroxyacetonitrile, HOCH₂CN; Zeng *et al.* 2018). In the RNA-world scenario, glycolonitrile is an intermediate species in the reductive homologation of HCN, and it is key in the construction of sugars (as e.g. glycolaldehyde and glyceraldehyde) needed for ribonucleotide assembly (see Patel *et al.* 2015). The advent of wide band, very sensitive instrumentation in radiotelescopes such as the broad-band Eight MIxer Receivers (EMIR) receivers at the Instituto de Radioastronomía Milimétrica (IRAM) 30m-diameter telescope or the Atacama Large Millimeter/Sub-millimeter Array (ALMA), together with the high degree of chemical complexity observed in the ISM, offer a unique opportunity to investigate to what extent the RNA-world chemical scheme proposed by Powner *et al.* (2009) and Patel *et al.* (2015) for the origin of life, can be extrapolated to the ISM.

In this paper, we explore whether precursors of pyrimidine ribonucleotides (according to the RNA-world chemical scheme of Powner *et al.* 2009 and Patel *et al.* 2015), are found in the ISM and star forming regions in our Galaxy. To do this, we use high sensitivity spectral surveys carried out toward a proto-Sun and a quiescent Giant Molecular Cloud in the Galactic Center to detect and determine the molecular abundances of species such as glyceraldehyde (CHOCHOHCH$_2$OH), dihydroxyacetone (DHA - CH$_2$OHCOCH$_2$OH), and 2-amino-oxazole (C$_3$H$_4$N$_2$O), key precursors in the primordial RNA-world chemical scheme. **These molecules remain undiscovered in space so far.** To our knowledge this is the first attempt to find 2-amino-oxazole in the ISM, a central molecule in the scheme proposed by Powner *et al.* (2009) and Patel *et al.* (2015). **Our high-sensitivity observations reveal that urea is a natural product of COM interstellar chemistry even without the intervention of star formation.** Other key species in the primordial RNA-world chemical scheme such as 2-amino-oxazole or dihydroxyacetone are not detected in our deep integrations, with upper limits to their abundance relative to molecular hydrogen of a few $\leq 10^{-11}$-$10^{-10}$. **The detection of these large COM precursors in the ISM would allow the field of Astrochemistry *to transition to Astrobiochemistry, i.e. to the study of biologically-relevant prebiotic compounds in space.***

## 2. Materials and Methods

### 2.1. Targeted molecular species.

In Fig. 1, we show a summary of **the chemical scheme** proposed for the synthesis of (pyrimidine) ribonucleotides, nucleobases (cytosine), and amino acids (glycine) within the RNA-world hypothesis for the origin of life (see Powner *et al.* 2009; Patel *et al.* 2015; Kitadai *et al.* 2018). **Another gas-phase route recently proposed by Rivilla *et al.* (2019b) for the formation of adenine in space has also been added to this diagram. Molecules detected already in the ISM and star-forming regions are indicated with solid-line boxes, while (so far) undetected molecules are shown within dotted-line boxes. The goal of this diagram is to visualize at** **once how many RNA-world prebiotic species have been detected in the ISM, and how they are related to each other according to the work by Powner et al. (2009), Patel et al. (2015), Kitadai et al. (2018) and Rivilla et al. (2019b). This diagram is not meant to imply any specific chemical reaction occurring in the ISM (because most of them take place in aqueous solution), but to identify the prebiotic species that are central in the primordial RNA-world theory for the origin of life, and that could be observed with current and future astronomical instrumentation.**

The molecular species for which spectroscopic information is available from either the Cologne Database for Molecular Spectroscopy (CDMS; Endres *et al.* 2016)[1] or from the Jet Propulsion Laboratory catalogue (JPL; Pickett *et al.* 1998)[2], are (see also Tables 1 and 2): hydrogen cyanide (HCN), formaldehyde ($H_2CO$), glycolonitrile ($HOCH_2CN$), glycolaldehyde ($CH_2(OH)CHO$), cyanamide ($NH_2CN$), glycolic acid ($CH_2OHCOOH$), methylamine ($CH_2NH$), urea ($NH_2CONH_2$), cyanomethanimine (E/Z-HNCHCN), aminoacetonitrile ($NH_2CH_2CN$), glyceraldehyde ($CHOCHOHCH_2OH$), dihydroxyacetone (DHA - $CH_2OHCOCH_2OH$), and glycine ($NH_2CH_2COOH$). The microwave spectrum of the $0^+$ and $0^-$ states of 2-amino-oxazole (currently unavailable in molecular spectroscopy databases) was recorded and analyzed by Møllendal & Konovalov (2010). These authors obtained the microwave spectrum of 2-amino-oxazole between 26.6 and 80 GHz by means of Stark-modulation spectroscopy. Due to the inversion of the amino group, there is a double-minimum potential that is manifested with the splitting of the lines into closely spaced doublets of comparable intensities, leading to two states designated as $0^+$ and $0^-$.

---

[1] http://www.astro.uni-koeln.de/cdms
[2] https://spec.jpl.nasa.gov/

From the fitting of the measured transitions with the Pickett's SPFIT/SPCAT program suite (Pickett 1991), two sets of experimental rotational parameters, including quartic centrifugal distortional constants, were obtained for the $0^+$ and $0^-$ states. **The spectroscopic information of these two states can be found in Table 3. For the details about the spectroscopic work of Møllendal & Konovalov (2010), please refer to their paper. Due to the lack of an entry for 2-amino-oxazole in molecular line catalogues such as JPL and CDMS, we have generated two transitions files (\*.cat files; see supplementary material) for the $0^+$ and $0^-$ states of 2-amino-oxazole, which have been implemented into the millimetre/sub-millimeter data analysis software MADCUBA (see Section 3.1). In this paper, we use these data** to search for 2-amino-oxazole toward two astronomical sources, a proto-Sun and a Giant Molecular Cloud located in the central region of the Milky Way.

*2.2. Astronomical sources.*

Our selected astronomical sources are the Solar-type protostar IRAS16293-2422 B and the quiescent Giant Molecular Cloud G+0.693-0.027 located in the Galactic Center. These two sources are among the richest ISM regions in COMs within our Galaxy, making them excellent targets for the search of new molecules in space (Requena-Torres *et al.* 2008; Jorgensen *et al.* 2016; Martin-Domenech *et al.* 2017; Zeng *et al.* 2018, 2019). In addition, they are located in two completely different environments: while IRAS16293-2422 B is part of a typical low-mass star-forming region in the Galactic disk just affected by the standard UV interstellar radiation field and standard cosmic-ray ionization rate (i.e. $G_0$=1 Habing or $1.6\times10^{-3}$ ergs cm$^{-2}$ s$^{-1}$ and $\xi$=3x10$^{-17}$ s$^{-1}$), G+0.693-0.027 is exposed to multiple highly-energetic phenomena taking place in the Galactic Center such as shock waves and enhanced cosmic rays. This will allow us to explore the effects of energy injection into the production of prebiotic COMs in the ISM. **In fact, the chemistry of COMs in the quiescent GMC G+0.693-0.027 can be used as a proxy of the COM chemistry experienced by the molecular cloud from which the proto-Solar nebula emerged, since the later likely underwent energetic processing by shock waves and enhanced cosmic rays as a result of (at least) one supernova explosion[3].**

IRAS16293-2422 B is a prototypical low-mass protostar that belongs to a multiple system in the ρ Ophiuchi star-forming cluster at a distance of ~141 pc (Dzib *et al.* 2018). With a luminosity of ≤3 Lo, this source represents the early stages in the formation of a Solar-type system (i.e. a proto-Sun). IRAS16293-2422 B is surrounded by a *hot corino*, a very dense (>$10^7$ cm-3; Jorgensen *et al.* 2016; Hernandez-Gomez *et al.* 2019), warm (>100 K; Ceccarelli *et al.* 2000; Bisschop *et al.* 2008), and compact condensation (~100 au; Zapata *et al.* 2013), which surrounds the nascent Solar-type protostar and where planets are expected to form. As a result of the high temperatures, the ice mantles from dust grains are sublimated, releasing all their organic content into the gas phase of the ISM. A wide variety of COMs have been reported toward this source (Caux *et al.* 2011; Jaber *et al.* 2014; Jorgensen *et al.* 2016; Rivilla *et al.* 2019a).

G+0.693-0.027 is a massive Giant Molecular Cloud (GMC) about 2.4 pc assuming a distance to the Galactic Centre of 8.34 kpc (Reid *et al.* 2014) across located within the Central Molecular Zone (CMZ, i.e. the central 500 pc) of the Milky Way (Morris & Serabyn 1996). Because of its location in the Galactic Center, the molecular gas in G+0.693-0.027 is affected by highly energetic phenomena such as low-velocity shocks and/or enhanced cosmic-ray ionization rates (Requena-Torres *et al.* 2006; Martin *et al.* 2008; Zeng *et al.* 2018). As a result, the physical conditions of G+0.693-0.027 are very different from those of IRAS16293-2422 B. Like the hot corino, G+0.693-0.027 presents high gas kinetic temperatures (between ∼50 K and ∼140 K; Guesten *et al.* 1985; Huettemeister *et al.* 1993; Krieger *et al.* 2017; Zeng *et al.* 2018). However, while in the hot corino gas and dust are thermally coupled due to the high densities of the gas (>$10^7$cm$^{-3}$), the H$_2$ gas density in G+0.693-0.027 is low (∼$10^4$ cm$^{-3}$), which yields also low dust temperatures (≤30 K; see Rodriguez-Fernandez *et al.* 2000, 2004). In addition, G+0.693-0.027 does not show any sign of star formation activity (in the form of either water masers, IR or mm continuum sources or UC HII regions; see e.g. Ginsburg *et al.* 2018). As found by Requena-Torres *et al.* (2008), G+0.693-0.027 is the quiescent GMCs in the Galactic Center with the highest level of chemical complexity, similar to that found in typical massive hot molecular cores such as SgrB2(N) (see e.g. Belloche *et al.* 2013). Large COMs such as propenal and propanal have been reported toward this cloud (Requena-Torres *et al.* 2008), as well as many N-bearing COMs including methylamine, methyl isocyanate, formamide or cyanomethanimine (Zeng *et al.* 2018; Rivilla *et al.* 2019b). G+0.693-0.027 is also the only GMC in the Galactic Center where the prebiotic phosphorus-bearing molecule PO has been found (Rivilla *et al.* 2018). All this makes G+0.693-0.027 not only the largest COM reservoir in our Galaxy (Requena-Torres *et al.* 2008), but also an excellent laboratory where to find prebiotic precursors of the RNA world.

---

[3] **The detection of short-lived radioactive species in meteorites suggests that the Sun's birthplace may have been a massive cluster affected by at least one supernova event (Adams 2010).**

We note that data from other hot molecular cores such as SgrB2 (N), Orion KL, W51 or NGC6334 I do exist in the ALMA data archive. However, the spectral surveys presented in this work likely suffer less from line confusion because: i) the molecular emission toward IRAS16293-2422 B, as observed with ALMA, has linewidths significantly narrower than those measured toward the aforementioned hot cores; and ii) the COM emission in the quiescent GMC G+0.693-0.027 is sub-thermally excited (derived excitation temperatures $T_{ex}$<15 K; see Section 3.1), which implies that only the lowest energy levels of the observed COMs are populated reducing the level of line confusion.

*2.3. Observations.*

Due to the different spatial scales of our selected astronomical sources, two different instruments have been used: the Atacama Large Millimeter/Sub-millimeter Array (ALMA) and the Instituto de Radioastronomia Milimetrica (IRAM) 30m diameter telescope. ALMA is an astronomical interferometer with 66 antennas that is located in the Chajnantor plateau at about 5000 m of elevation. ALMA operates at wavelengths from 3 to 0.3 mm. The IRAM 30m telescope is a single-dish antenna located in Pico Veleta (Sierra Nevada, Spain). It operates at the 3, 2, 1 and 0.9 mm atmospheric bands that are transparent to external (sub)millimeter radiation coming from space. While single-dish antennas are typically used to observe extended astronomical sources in the sky such as the quiescent (and pc-scale[4]) GMC G+0.693-0.027 in the Galactic Center, interferometers such as ALMA are needed to image with a high level of detail the molecular emission arising from the more compact Sun-like protostars such as IRAS16293-2422 B (with a typical size of ~100 au[5]).

---

[4] 1 pc (or parsec) corresponds to $2x10^5$ astronomical units (au) or $3x10^{13}$ km.

For the observations of the IRAS16293-2422 B hot corino, we have used several datasets obtained with ALMA in Bands 3, 4, 6, 7 and 8, and which are publicly available from the ALMA data archive (see e.g. Jørgensen *et al.* 2016; Martin-Domenech *et al.* 2017). The spectral line data cubes used for the identification of new molecular species and for the determination of their abundances, were obtained by subtracting the continuum emission in the uv-plane before doing the imaging using line-free channels from the observed spectra. The standard ALMA calibration scripts and the Common Astronomy Software Applications package (CASA; McMullin *et al.* 2007)[6] were used for data calibration and imaging. This dataset covers a total bandwidth of ~14GHz split into multiple spectral ranges between 86.5 GHz and 353.5 GHz. The beam sizes range between 1.42" and 1.85" (equivalent to 200-260 au), while the spectral resolutions lie between 61-282 kHz corresponding to velocity resolutions of 0.1-0.9 km s$^{-1}$. For the analysis, we have considered a spatial circular support of 1.6" diameter to extract the spectra from each spectral window toward the position of IRAS16293-2422 B: $\alpha$(J2000.0)=16h32m22.61s , $\delta$(J2000.0)=-24∘28'32".4400. Note that the beam sizes of the dataset are larger than the size of source B (0.5"; see Zapata *et al.* 2013; Jørgensen *et al.* 2016; Lykke *et al.* 2017; Martín-Doménech *et al.* 2017) and hence the flux measured within our circular support of 1.6" diameter contains all the flux independently of the beam. The beam sizes ≤1.9" are sufficient to resolve source B from its companion source A in the IRAS16293-2422 binary and therefore, the typical linewidths of the emission from IRAS16293-2422 B are ≤2 km s$^{-1}$.

---

[5] 1 astronomical unit (au) corresponds to $1.5 \times 10^8$ km
[6] https://casa.nrao.edu

The observations of the quiescent GMC G+0.693-0.027 were carried out in April **and August 2019** with the IRAM 30m diameter telescope located at Pico Veleta (Spain). The single-pointing coordinates of the source are α(J2000.0)= 17h 47m 22s and δ(J2000.0)= -28∘ 21' 27". The position switching mode was used in for observations and the coordinates of the reference position were α(J2000.0)= 17h 46m 23.01s and δ(J2000.0)= -28∘ 16' 37.3". By using the broadband Eight MIxer Receivers (EMIR), we observed the full 3 mm band (72-116 GHz) and partially covered the 2 mm and 1 mm **bands (spectral ranges from 125 to 172 GHz, and from 208 to 238 GHz).** The FTS200 mode of the backends was used, which provided spectral resolutions of ~200 kHz, equivalent to 0.3-0.8 kms$^{-1}$ between 72 and 238 GHz. **To increase the signal-to-noise ratio in some parts of the survey, we smoothed the final reduced spectra to a frequency resolution of ~400 kHz (i.e. 0.6-1.6 km s$^{-1}$ between 72 and 238 GHz).** This velocity resolution is sufficient to resolve the typical linewidths of the molecular emission toward this source (~20 km s$^{-1}$; Zeng *et al.* 2018). The half-power beam widths (HPBW) of the telescopes were in the range ~9"- 35" (equivalent to $8x10^4$-$3x10^5$ au at the distance of the source of 8.5 kpc). Typical system temperatures, Tsys, ranged between 100 K and 300 K in the 3mm, 2mm and 1mm atmospheric bands, respectively. The line intensity of our spectra is given in antenna temperature corrected for atmospheric attenuation, $T_A^*$, as the molecular emission toward G+0.693 is extended over the beam of the radiotelescope (Requena-Torres *et al.* 2006; Martin *et al.* 2008; Rivilla *et al.* 2018).

We finally note that, for some species such as glycolaldehyde, we have used previous, less sensitive surveys on G+0.693-0.027 carried out with the IRAM 30m telescope and the Green Bank Telescope (GBT). The details of these observations are described in detail in Zeng *et al.* (2018) and Rivilla *et al.* (2019b).

### 3. Results

*3.1. Molecular line identification, excitation temperatures and column densities.*

The line identification and analysis were performed using the MADCUBA (Madrid Data Cube Analysis on ImageJ) software package developed at the Center for Astrobiology (CAB) in Spain (see Rivilla *et al.* 2016; Martin *et al.* 2019). This software uses the spectroscopic data available from databases such as CDMS and JPL to carry out the radiative transfer of a given species and for a given set of physical parameters such as column density ($N_{tot}$), excitation temperature ($T_{ex}$), central radial velocity ($V_{LSR}$), and linewidth ($\Delta v$). The MADCUBA-SLIM (Spectral Line Identification and Modeling) tool then produces synthetic spectra for each line profile by considering Local Thermodynamical Equilibrium (LTE) and line opacity effects. The parameters are initially adjusted manually to match the synthetic spectra to the observed line profiles and then the MADCUBA-AUTOFIT tool is finally employed to provide the best non-linear least-squared fit using the Levenberg-Marquardt algorithm. In Tables 1 and 2 (see columns 3-6), we report the derived molecular column densities Ntot (and their $3\sigma$ upper limits in case of non-detections), excitation temperatures $T_{ex}$, central radial velocities ($V_{LSR}$), and linewidths ($\Delta v$), obtained for our targeted molecules using this method. **In Tables 6 and 7, we list the detected**

**COM rotational transitions together with their quantum numbers, frequencies, line intensities at 300 K [$\log_{10}I(300K)$], energies of the lower level ($E_l$), linewidths ($\Delta v$), radial velocities ($V_{LSR}$) and integrated intensities of the lines (Area). In these Tables, we also include the transitions that have been used to estimate the upper limits to the column densities and abundances of the COMs that have not been detected in our observations.**

We note that for the typical densities and linear scales imaged with ALMA toward the hot corino IRAS16293-2422, LTE is a good approximation (see Jørgensen *et al.* 2016). **For the quiescent GMC G+0.693-0.027, the low $H_2$ densities of the gas in this region (~$10^4$ cm$^{-3}$; Rodriguez-Fernandez *et al.* 2004) could yield weak maser amplification in the molecular excitation of large COMs such as those studied here. However, those non-LTE effects appear at centimeter wavelengths (especially at frequencies <30 GHz) rather than in the millimeter range (see Faure *et al.* 2014, 2018). The excitation temperatures $T_{ex}$ derived from the COM emission in this source are <15 K (Table 2 and Requena-Torres *et al.* 2006, 2008; Zeng *et al.* 2018), which reflects the fact that these molecules are sub-thermally excited as a result of the low $H_2$ densities of the gas in this region. Since collisional rate coefficients with molecular $H_2$ are currently not available for these large COMs (collisional rate coefficients have been calculated only for methyl formate, $CH_3OCHO$, and methanimine, $CH_2NH$; Faure *et al.* 2014, 2018), LTE analysis is the only method available to determine the excitation conditions of these COMs in the quiescent GMC G+0.693-0.027.**

To calculate the molecular abundances relative to $H_2$, we consider an $H_2$ column density of $2.8 \times 10^{25}$ cm$^{-2}$ for IRAS16293-2422 as inferred by Martin-Domenech *et al.* (2017), and of $1.35 \times 10^{23}$ cm$^{-2}$ for G+0.693-0.027 as derived by Martin *et al.* (2008). The abundances (or upper limits) of all species are provided in column 7 of Tables 1 and 2.

**From Figures 2-3 and Tables 6 and 7, we find clear detections of HC$^{15}$N, H$_2$C$^{18}$O/H$_2^{13}$CO, glycolonitrile, glycolaldehyde, cyanamide, methanimine and E-/Z-cyanomethanimine within the ALMA and IRAM 30m datasets toward either or both astronomical sources. In addition, Figure 4 and Table 7 report the detection of urea toward the quiescent GMC G+0.693-0.027. This represents the first detection of this pre-biotic molecule outside a star-forming region (urea has recently been found for the first time toward the SgrB2 N massive hot core; see Belloche *et al.* 2019). Table 7 lists the unblended transitions of urea measured toward G+0.693-0.027 (8 in total, but three are doublets). The observed intensities of the rest of transitions are consistent with our MADCUBA simulations because they are either blended with other molecular lines or fall below the rms noise level of our observations (typically of ~3 mK in a velocity resolution of ~1 km s$^{-1}$). Note that the transitions at 102.864 GHz and 102.767 GHz (shown in Figure 4) have been identified by Belloche *et al.* (2019) as clean from any molecular contaminant. In addition, since the molecular excitation temperature in this source is $T_{ex}$<15 K, we do not expect to detect many more low-excitation transitions of urea in the observed spectra besides those reported in Figure 4 and Table 7. The spectrum of urea can be fitted well by using an excitation temperature $T_{ex}$=8 K (consistent with the $T_{ex}$ measured for other COMs in this source; see Requena-Torres *et al.* 2006, 2008; Zeng *et al.* 2018), and a column density of $(6.3\pm0.1) \times 10^{12}$ cm$^{-2}$ (see Figure 4 and**

**Table 2). This implies a urea abundance of ~5x10⁻¹¹ toward the quiescent GMC G+0.693-0.027.** Glycolaldehyde, cyanamide, methanimine and E-/Z-cyanomethanimine have been already reported by our group toward this source (see Table 2 and Requena-Torres *et al.* 2008; Zeng *et al.* 2018; Rivilla et al. 2019b).

Our simulations with MADCUBA reveal that the isotopologues $HC^{15}N$, $H_2C^{18}O$ and $H_2^{13}CO$ are optically thin toward IRAS16293-2422 B and G+0.693-0.027. Therefore, large optical depth effects do not likely affect their column densities reported in Tables 1 and 2.

Toward the proto-Sun IRAS16293-2422 B, glycolaldehyde, cyanamide and methanimine had been previously found by other groups; our derived column densities and excitation temperatures agree within factors of 10 and 2, respectively, with those inferred elsewhere (see Bisschop *et al.* 2008; Jorgensen *et al.* 2016; Coutens *et al.* 2018; Ligterink *et al.* 2018; Persson *et al.* 2018, Rivilla *et al.* 2019a). For glycolonitrile and cyanamide, these species are fitted better using two temperature components along the line-of-sight, a hot (T>100 K) and a cold (T<30 K) component. These components are physically associated with the hot corino and the colder envelope around IRAS16293-2422 B respectively, although in reality they may reflect a decreasing temperature gradient within the envelope of this source for increasing radii from the central protostar (see also Zeng *et al.* 2019 for a detailed discussion on these two temperature components). In Table 1, however, we only present the results for the hot component associated with the hot corino. **Urea has not been detected toward the IRAS16293-2422 B hot corino, with a derived upper limit to its abundance of ≤ 2.3x10⁻¹¹.**

The species glycolic acid, amino acetonitrile, glyceraldehyde, dihydroxyacetone, glycine and 2-amino-oxazole, have not been detected, so far, in any of the sources (see the upper limits to their derived column densities in column 6 of Tables 1 and 2). However, the unprecedented sensitivity of these observations provides stringent constraints to the upper limits of the abundances measured for these prebiotic COMs. **Particularly stringent are the ones obtained toward G+0.693-0.027 thanks to our new high-sensitivity IRAM 30m observations. These upper limits range between $\leq 0.4\text{-}17 \times 10^{-11}$ for glycolic acid, $\leq 0.4\text{-}5 \times 10^{-11}$ for amino acetonitrile, $\leq 5\text{-}9 \times 10^{-11}$ for glyceraldehyde, $\leq 0.5\text{-}4 \times 10^{-10}$ for dihydroxyacetone, $\leq 2\text{-}4 \times 10^{-10}$ for glycine, and $\leq 1\text{-}8 \times 10^{-11}$ for 2-amino-oxazole.**

*3.2. Column density ratios of sugar-like species and N-bearing prebiotic COMs.*

In Tables 4 and 5, we report the column density ratios of sugar-like species (glycolaldehyde, glycolic acid, glyceraldehyde and dihydroxyacetone) with respect to formaldehyde ($H_2CO$), and between N-bearing prebiotic species (glycolonitrile, cyanamide, methanimine, urea, amino acetonitrile, cyanomethanimine, glycine and 2-amino-oxazole) with respect to hydrogen cyanide (HCN). In order to calculate these ratios, we first need to derive the total column densities of $H_2CO$ and HCN toward IRAS16293-2422 B and G+0.693-0.027. Formaldehyde ($H_2CO$) emission is optically thick toward both sources. However, we can use the optically thin isotopologues $H_2C^{18}O$ toward IRAS16293-2422 B and $H_2^{13}CO$ toward G+0.693-0.027. For IRAS16293-2422 B we have assumed a $^{16}O/^{18}O$ isotopic ratio of ~805 corresponding to that measured by Persson *et al.* (2018), while for G+0.693-0.027 we have considered a $^{12}C/^{13}C$ isotopic ratio of ~20 as found in the Galactic Center (Wilson & Rood 1994). By using these isotopic ratios, we derive $H_2CO$ total column densities of $2.8x10^{18}$ cm$^{-2}$ toward IRAS16293-2422 B and of $4x10^{14}$ cm$^{-2}$ toward G+0.693-0.027. These $H_2CO$ total column densities are similar to those estimated by Persson *et al.* (2018) for IRAS16293-2422 B (of $2x10^{18}$ cm$^{-2}$) and by Requena-Torres *et al.* (2006, 2008) toward G+0.693-0.027 (of $6.4x10^{14}$ cm$^{-2}$).

HCN emission is also optically thick toward both sources. In order to determine its column density, we use the $^{15}N$ isotopologue of HCN (HC$^{15}$N; see Tables 1 and 2, and Tables 6 and 7). The HC$^{15}$N line observed with ALMA toward IRAS16293-2422 B shows an inverse P-Cygni profile with bright blue-shifted emission and a red-shifted component seen in absorption. This type of profile has already been observed toward this source in the spectra from other molecular species such as $CH_3OCHO$, $CH_3OH$ or HCO (see Pineda *et al.* 2012; Rivilla *et al.* 2019a). The HC$^{15}$N line profile was fitted with the MADCUBA-AUTOFIT tool by assuming a source size of 0.5" and a Tex=150 K for the hot corino and by modeling the continuum emission using a modified black body with $T_c$=180 K, dust opacity index $\beta$=0 and $\tau$(94 GHz)=2.1 (see also Rivilla *et al.* 2019a). For G+0.693-0.027, the HC$^{15}$N line only shows emission and it has been modeled without considering any continuum source. As isotopic ratios, we assume a $^{14}N/^{15}N$ ratio of ~163-190 for IRAS16293-2422 B, i.e. the same as the one measured by Wampfler *et al.* (2014) toward IRAS16293-2422 A. For G+0.693-0.027, we consider the lower limit to the $^{14}N/^{15}N$ ratio >600 obtained by Wilson & Rood (1994). The inferred total column densities of HCN are therefore $(5.7-6.7)x10^{16}$ cm$^{-2}$ for IRAS16293-2422 B and $>6.6x10^{15}$ cm$^{-2}$ for G+0.693-0.027.

From Table 4, we obtain very stringent constraints to the abundance of sugar-like species relative to $H_2CO$: ~1-2% for glycolaldehyde, ≤0.004% for glycolic acid, ≤0.06% for glyceraldehyde and ≤0.4% for dihydroxyacetone in IRAS16293-2422 B; ~8% for glycolaldehyde, ≤6% for glycolic acid, ≤3% for glyceraldehyde and ≤1.7% for dihydroxyacetone in G+0.693-0.027. **Interestingly, glycolaldehyde, the only sugar-like species detected, presents column density ratios of the same order toward IRAS16293-2422 B and G+0.693-0.027 (0.01-0.02 and 0.08, respectively), although it seems to be produced more efficiently in the quiescent Galactic Center GMC.** One possible explanation is that the energetic processing of the gas and the icy mantles of dust grains by shocks and/or cosmic rays in the Galactic Center, favors the formation of sugar-like species such as glycolaldehyde. **Indeed, it has been proposed that glycolaldehyde formation is boosted by ice irradiation (Biver *et al.* 2015).** This irradiation can either be in the form of UV photons, X-rays or cosmic rays, **since all types of irradiation have the same effect on the final chemical composition of the ices** (see de Marcellus *et al.* 2014; Muñoz Caro *et al.* 2014; Meinert *et al.* 2016; Ciaravella *et al.* 2019). Although G+0.693-0.027 is not close to any obvious source of radiation, this cloud is embedded in an environment where the cosmic-ray ionization rate is enhanced by factors 100-1000 (Goto *et al.* 2013). This yields an also enhanced radiation field of comic-ray-induced, secondary UV-photons (by the same factors), which has a clear effect on the observed chemistry (Harada *et al.* 2015; Zeng *et al.* 2018). Therefore, the data presented here suggests that the formation of sugar-like species is boosted by irradiation in the ISM.

From Table 5, we find that glycolonitrile is about 3% the amount of HCN measured in IRAS16293-2422 B, while it is undetected in the quiescent GMC G+0.693-0.027 (≤0.1% the amount of HCN). **On the contrary, cyanamide and cyanomethanimine are more abundant with respect to HCN in G+0.693-0.027 than in IRAS16293-2422 B (by factors of 5 and ≥1.5-2.5, respectively). The rest of N-bearing species show similar ratios toward both sources. Important upper limits to the abundance ratios with respect to HCN are those obtained for glycine (≤0.9%) and 2-amino-oxazole (≤0.2%; see Table 5).**

**The different trends observed in the abundance ratios with respect to HCN for glycolonitrile (HOCH$_2$CN), and for cyanamide (NH$_2$CN) and cyanomethanimine (HNCHCN), may be related to the different chemistries found in these sources.** G+0.693-0.027 is a turbulent GMC affected by large-scale shocks widespread across the Galactic Center, and by an enhanced cosmic ray ionization rate (Harada *et al.* 2015; Goto *et al.* 2013; Zeng *et al.* 2018). The energetic processing of the molecular gas in this region favors the formation of unsaturated (hydrogen-poor) molecules such as cyanamide and cyanomethanimine. In contrast, the chemistry in IRAS16293-2422 B is dominated by protostellar heating. During the warm-up phase, saturated (hydrogen-rich) species such as glycolonitrile have enough time to form on the surface of dust grains (see the chemical modeling of glycolonitrile for IRAS16293-2422 in Zeng *et al.* 2018) and are thermally sublimated and incorporated into the gas phase once the temperature of the dust reaches 100 K (i.e. the desorption temperature for water; see e.g. Collings *et al.* 2004).

## 4. Discussion

*4.1. Comparison to previous searches of large pre-biotic COMs: glycine, dihydroxyacetone and glyceraldehyde*

Because of its simplicity, glycine has been the only amino acid extensively searched for in the ISM. Deep searches have been carried out toward a variety of astronomical objects such as massive hot cores (as e.g. in SgrB2, Orion KL, W51 e1/e2; Kuan *et al.* 2003; Belloche *et al.* 2003, 2008), low-mass hot corinos (as in IRAS16293-2422; Ceccarelli *et al.* 2000) and pre-stellar cores (L1544; Jimenez-Serra *et al.* 2016), but all of them have been unsuccessful. The upper limit derived in this work for glycine toward the hot corino in IRAS16293-2422 B ($\leq 2.1 \times 10^{-10}$; Table 1) is a factor $\geq 30$ lower than that previously reported by Ceccarelli *et al.* (2000; of $\leq 7 \times 10^{-9}$). This implies that the amount of glycine formed in the ices during the warming-up of the protostellar envelope is $\leq 0.0003\%$ the abundance of water, assuming all ice content has been released into the gas phase in the hot corino evaporation phase and considering a water abundance of $7.25 \times 10^{-5}$ with respect to $H_2$ (Whittet & Duley 1991). This low relative abundance of glycine with respect to water is a factor of $\sim 50$ lower than that obtained in UV-irradiated interstellar ice analogs (glycine abundance of $\sim 10^{-4}$ with respect to water, or $\sim 10^{-8}$ with respect to $H_2$; Muñoz Caro *et al.* 2002), and it is consistent with the abundance range inferred by ROSINA in comet 67P/Churyumov-Gerasimenko (between 0 and 0.0025 with respect to water; see Altwegg *et al.* 2016). Irradiation can yield a wide variety of amino acids in ices under interstellar conditions (Muñoz Caro *et al.* 2002; Bernstein *et al.* 2002). However, the upper limit to the abundance of glycine measured toward the quiescent GMC G+0.693-0.027 gives an abundance of $\leq 4 \times 10^{-10}$ or

≤0.0006% with respect to water (Table 2). **As mentioned in Section 3.2, the GMC G+0.693-0.027 is located in the Galactic Center and it is affected by enhanced cosmic ray ionization rates. Since both UV and cosmic ray irradiation yield a similar chemistry and COM composition in interstellar ices (see e.g. Muñoz Caro *et al.* 2014), our measured upper limit to the abundance of glycine suggests that ice irradiation may not produce glycine efficiently in the ISM. Alternatively, this molecule could form on irradiated interstellar ices but, once injected into the gas phase by proto-stellar feedback, glycine could be destroyed quickly by ions and radicals (Garrod 2013; Suzuki et al. 2018).**

In contrast to glycine, dihydroxyacetone and glyceraldehyde have only been searched for toward the SgrB2 N-LMH massive hot core (Widicus-Weaver *et al.* 2005; Hollis *et al.* 2004; **Apponi *et al.* 2006**). For dihydroxyacetone, Widicus-Weaver *et al.* (2005) reported a tentative detection of this large prebiotic COM through measurements of nine possible rotational transitions of this molecule with the Caltech Sub-millimeter Observatory (CSO) telescope. The inferred abundance of dihydroxyacetone in this source is $1.2 \times 10^{-9}$. This tentative detection has never been confirmed **(see Apponi *et al.* 2006)**, which is in line with our findings. The derived upper limits to the abundance of dihydroxyacetone in IRAS16293-2422 B and G+0.693-0.027 range from $\leq 1.8$-$3.9 \times 10^{-10}$, i.e. factors 3-7 lower than the abundance measured in SgrB2 N-LMH by Widicus-Weaver *et al.* (2005). For glyceraldehyde, Hollis *et al.* (2004) reported just upper limits to the abundance of this molecule, which is consistent with our non-detection (upper limit to the abundance of glyceraldehyde of $\leq 2.4$-$5.7 \times 10^{-11}$; see Tables 1 and 2). This implies that the so called "formose reaction" by which formaldehyde polymerizes on interstellar dust grains into glycolaldehyde first and then into glyceraldehyde and into more complex sugars (Larralde *et al.* 1995), does not seem to operate in the ISM.

### 4.2. Comparison to the COM content in comets

**One possible scenario for the appearance of prebiotic material on a young Earth is the delivery of such material by the impact of comets and/or meteorites on the Earth's surface. It is currently believed that the COM content in comets is directly linked to the pristine chemical composition of the Solar Nebula, i.e. to the chemical composition developed during the first stages of low-mass star formation. This is supported by the high deuterium enrichment measured in comets, which can only be explained if the composition of their icy surfaces was set in the presolar cloud at large distances from the Sun (Ceccarelli *et al.* 2014; Altwegg *et al.* 2015). Therefore, the comparison between the prebiotic COM content in comets and in the ISM can provide insight into how life originated on Earth.**

**HCN and $H_2CO$ are among the most abundant species in comets such as C/2014 Q2 Lovejoy (Biver *et al.* 2015) and 67P/Churyumov-Gerasimenko (Goesmann *et al.* 2015). We have thus calculated the COM ratios (with respect to HCN and $H_2CO$) available in the literature and compared them with the ones provided in Tables 4 and 5. For sugar-like species, the glycolaldehyde/$H_2CO$ abundance ratio is ~0.05 in comet C/2014 Q2 Lovejoy, as inferred from their relative abundances to water reported in Table 1 of Biver *et al.* (2015). This ratio falls within the range of the glycolaldehyde/$H_2CO$ ratios measured toward IRAS16293-2422 B (0.01-0.02) and G+0.693-0.027 (0.08; see Table 4). The slightly higher abundance of glycolaldehyde in comets with respect to the hot corino stage (see also Table 4** and Figure 2 in Biver *et al.* 2015), can be attributed to the synthesis of this sugar-like species through grain-surface reactions and subsequent ice irradiation in the solar nebula. This idea is supported by the higher glycolaldehyde/$H_2CO$ ratio measured toward G+0.693-0.027, which is exposed to the enhanced cosmic-ray ionization rates of the Galactic Center.

For N-bearing prebiotic species, the abundance ratio of glycine/HCN measured in comet 67P/Churyumov-Gerasimenko is ~0.003, which is consistent with the upper limits obtained toward the two sources (≤0.1 for IRAS16293-2422 B and ≤0.009 for G+0.693-0.027). This suggests that the COM material in comets could have an ISM origin. Unfortunately, glycine is the only N-bearing molecule from our list of prebiotic species for which comet data are available. Additional remote observations of species such as glycolonitrile, cyanamide or methanimine toward comets are needed to better characterize the link between the COM prebiotic content in these objects and that of the ISM.

*4.3 Urea as a common prebiotic product of ISM chemistry*

Urea is a key molecule in animal life because it plays an important role in the metabolism of N-containing compounds. Until recently, the detection of urea had remained elusive. However, Belloche *et al.* (2019) have reported recently the first detection of this prebiotic species in the ISM toward the hot molecular envelope around a massive(s) protostar (the SgrB2 N hot core; Belloche *et al.* 2019). Our high-sensitivity observations carried out with the IRAM 30m telescope toward the quiescent GMC G+0.693-0.027, not only provide the second detection of urea in space (with a derived abundance of $4.7 \times 10^{-11}$), but it also shows that this molecule can be formed in the ISM in regions unaffected by star formation.

Belloche *et al.* (2019) does not provide any absolute abundance of urea in their work and therefore a direct comparison with the urea abundance estimated toward G+0.693-0.027 cannot be performed. However, we can check the relative abundance of this prebiotic molecule with respect to other chemically-related species such as methyl isocyanate ($CH_3NCO$) and formamide ($NH_2CHO$; see Belloche *et al.* 2019). By using the column densities of methyl isocyanate and formamide derived toward G+0.693-0.027 by Zeng *et al.* (2018), we infer that the Urea/$CH_3NCO$ and Urea/$NH_2CHO$ column density ratios are ~0.1 and ~0.01 respectively. From Table 4 of Belloche *et al.* (2019), we can calculate the same column density ratios toward the massive hot core SgrB2 N1, which gives ~0.11 for Urea/$CH_3NCO$ and ~0.009 for Urea/$NH_2CHO$. These ratios are very similar to the ones found in G+0.693-0.027. This suggests that the chemistry of urea likely follows similar formation/destruction pathways both in hot cores and in Galactic Center quiescent Giant Molecular Clouds.

Belloche *et al.* (2019) have proposed that urea is formed directly from formamide on the surface of dust grains. The process starts with the abstraction of H from formamide ($NH_2CHO$), which yields the intermediate species $NH_2CO$ that is then free to react with $NH_2$ to form urea ($NH_2CONH_2$). This scenario is supported by recent laboratory experiments of UV irradiation of $CH_4$:HNCO ice mixtures (Ligterink *et al.* 2018), which produce the intermediate radical $NH_2CO$ in high amounts. Alternatively, urea could be formed in the ISM via the gas phase reaction $NH_2OH_2^+ + NH_2CHO$ although, as recently shown by Jeanvoine & Spetia (2019), this reaction does not proceed under interstellar conditions.

From all this, urea formation seems to be dependent on the production of formamide. According to our results, the formation yield of urea is 1% that of formamide. The production of formamide in the ISM has been the subject of debate in recent years. Some groups have proposed that formamide is formed on the surface of dust grains (e.g. Fedoseev *et al.* 2016; Dulieu *et al.* 2019), while others suggest that formamide is produced by pure gas phase reactions such as $NH_2 + H_2CO \rightarrow NH_2CHO + H$ (e.g. Skouteris *et al.* 2017). Alternatively, Quénard et al. (2018) have proposed that the production of formamide may be a combination of the two processes (gas-phase and grain surfaces), each of them dominating at different physical regimes. For the physical conditions found in hot molecular cores and in the Galactic Center quiescent GMCs, grain surface formation dominates, which explains why the Urea/$NH_2CHO$ column density ratios measured toward SgrB2 N1 and G+0.693-0.027 are very similar. Further laboratory experiments and theoretical calculations are needed to expand our understanding on the chemistry of urea in the ISM.

*4.4. The missing link: 2-amino-oxazole*

In the RNA-world chemical scheme, 2-amino-oxazole plays a central role in the formation of pyrimidine ribonucleotides (see Powner *et al.* 2009 and Patel *et al.* 2015, and Fig. 1). In this work, we have presented the first attempt to detect this large COM in the ISM. The derived upper limits to its abundance (for both states $0^+$ and $0^-$) toward IRAS16293-2422 B and G+0.693-0.027 are $\leq(1\text{-}8)\text{x}10^{-11}$ with respect to $H_2$ or $\leq 0.002\text{-}0.02$ with respect to HCN. The lack of detection may be due to two reasons: i) the transitions covered within our observations may not have been the most favorable for a detection experiment; or ii) the sensitivity of the observations is not high enough for a detection experiment of 2-amino-oxazole in the ISM.

To check case i), we have simulated the spectrum of 2-amino-oxazol (for both states 0+ and 0-) using MADCUBA and considering the physical conditions and the upper limits to the abundance of 2-amino-oxazole measured in IRAS16293-2422 B and G+0.693-0.027. When the excitation temperature of the gas ($T_{ex}$) is low (of ~10 K, as in G+0.693-0.027) the brightest lines of 2-amino-oxazole $0^+$ and $0^-$ appear at frequencies 87273/87300 MHz and 105921/105952 MHz (i.e. the $5_{5,K2}\text{-}4_{4,K2'}$ and $6_{6,K2}\text{-}5_{5,K2'}$ transitions of the $0^-/0^+$ states, respectively). These transitions were covered with the IRAM 30m telescope toward G+0.693-0.027. However, for an excitation temperature of $T_{ex}$=150 K (as assumed for most COM material in IRAS16293-2422), the brightest lines of 2-amino-oxazole should be found at 256585/256670 MHz (the $16_{13,K2}\text{-}15_{12,K2'}$ transitions of the $0^-/0^+$ states, respectively). These transitions are not included in the ALMA dataset used here.

To test case ii), the upper limits to the abundance of 2-amino-oxazole are still high since the lowest values measured in the ISM for the abundance of COMs are of the order of ~$10^{-12}$-$10^{-13}$ (see e.g. Jimenez-Serra *et al.* 2016). Therefore, while for G+0.693-0.027 the non-detection of 2-amino-oxazole may have been related to the lack of sensitivity, for IRAS16293-2422, it may have been due to both, poor sensitivity and lack of bright enough transitions covered within the observations.

*4.5. Evolution of Astrochemistry and prebiotic chemistry: Towards Astrobiochemistry in the ISM*

The first experiments of prebiotic chemistry used simple molecules such as $H_2O$, $CH_4$, $NH_3$ or HCN as precursors of proteinogenic amino acids (such as glycine and alanine) and of simple nucleobases (such as adenine; Miller 1953; Oró 1961). The use of these simple species were motivated by their early discovery in space (see e.g. Cheung *et al.* 1969; Snyder & Buhl 1971). However, subsequent studies showed that early Earth's atmosphere may have been richer in other gases such as $CO_2$ or $N_2$ rather than $H_2O$ or $NH_3$; and in addition, although amino acids are formed in the Miller-Urey experiment, they cannot be assembled spontaneously into polypeptide chains (i.e. proteins) without the intervention of RNA, which led to researchers to propose the RNA-world scenario for the origin of life (see review by e.g. Ruiz-Mirazo *et al.* 2013).

The chemical scheme of Powner *et al.* (2009) uses more complex molecules such as cyanamide or glycolaldehyde as precursors of the RNA formation process. This may have been a conceptual problem in the 70's and 80's, because these molecules (believed to be injected into a young Earth through the impact of comets and meteorites on the Earth's surface) were unknown in the ISM. In the past decade, however, Astrochemistry has been extremely successful at detecting new COMs in space, which demonstrates the increasingly high level of chemical complexity found in the ISM (Herbst & van Dishoeck 2009). Indeed, COMs such as cyanamide and glycolaldehyde are nowadays routinely detected across multiple environments in the ISM (e.g. low-mass and high-mass protostars, Galactic Center GMCs; Coutens *et al.* 2018; Jorgensen *et al.* 2012; Requena-Torres *et al.* 2008; Zeng *et al.* 2018). In addition, our recent discoveries of glycolonitrile and Z-cyanomethanimine (a precursor of adenine) toward the hot corino IRAS16293-2422 B and the quiescent GMC G+0.693-0.027 in the Galactic center respectively (Zeng *et al.* 2019; Rivilla *et al.* 2019b), **as well as the detection of urea (Belloche *et al.* 2019, and this work),** not only demonstrates the wide variety of chemical compounds that can be formed under interstellar conditions, **but it also shows that ISM chemistry, by itself, tends to form precursors that are essential for the synthesis process of RNA. This means that if these compounds were delivered to the surface of a young planet by the impact of comets and/or meteorites, they could trigger downstream in the chemical scheme of Patel *et al.* (2015) the production chain of ribonucleotides in aqueous solution without having to invoke the first reactions. Our analysis below shows that the reaction formation/destruction processes of these prebiotic species are very different in space from those occurring in aqueous solutions. But our intention in this work is to show that, if they were injected early into the reaction network of the primordial RNA-world chemical scheme of Patel *et al.* (2015), they would facilitate the subsequent production of ribonucleotides. Indeed, experiments of simulated meteorite impacts of prebiotic species such as glycolaldehyde mixed with clay, reveal that these COM molecules not only survive the impact, but they also react forming new biologically relevant molecules (McCaffrey *et al.* 2014).**

**The different reaction processes involved in the formation/destruction of prebiotic species in the ISM and in aqueous solution, become apparent if one pays attention to the molecular abundances derived for these precursors toward IRAS16293-2422 B and G+0.693-0.027.**

**Tables 1 and 2 show that** glycolonitrile (a key precursor of glycolaldehyde in the chemical scheme proposed by Patel *et al.* 2015), is at least one order of magnitude less abundant than glycolaldehyde. In fact, glycolonitrile is not even detected toward G+0.693-0.027, which contrasts the relatively high abundance of glycolaldehyde measured toward this source (~3x10$^{-10}$). This rules out glycolonitrile as precursor of glycolaldehyde in space.

**The formation of glycolaldehyde in the ISM has remained a mystery for a long time since this species tends to be more abundant in regions with temperatures colder than those found in hot cores. This is the case of the line-of-sight toward the hot molecular core SgrB2 N, where glycolaldehyde presents a higher abundance in the molecular envelope than in the hot core (Hollis *et al.* 2000). One possibility is that glycolaldehyde is formed in the gas phase from COM precursors such as methanol or formic acid (in their protonated forms; Laas *et al.* 2011) or ethanol (Skouteris *et al.* 2018). However, recent chemical modeling and observations suggest that glycolaldehyde may form more efficiently on the surface of dust grains via either the dimerization of HCO followed by hydrogenation, or the reaction between the radicals HCO and CH$_2$OH (Woods *et al.* 2013; Rivilla *et al.* 2017, 2019a; Coutens *et al.* 2018). The observed trend for the abundance ratio between ethylene glycol/glycolaldehye to increase with increasing luminosity supports the grain surface formation route for glycolaldehyde (Coutens *et al.* 2018; Rivilla *et al.* 2019a).**

For cyanamide, this molecule is a factor of 10 less abundant than methanimine (at least toward IRAS16293-2422 B, a proto-Sun), which suggests that the latter species may be the actual precursor of larger amine COMs (Coutens *et al.* 2018).

From Tables 1 and 2 it is clear that the derived upper limits to the abundance of key species in the chemical scheme of the RNA-world scenario such as 2-amino-oxazole, glyceraldehyde or dihydroxyacetone, are still high (between $10^{-11}$ and $10^{-10}$). These upper limits therefore do not allow us to rule out the possibility that these species are not present in the ISM. Particularly important is the molecule 2-amino-oxazole, which is a bottleneck in the pyrimidine nucleotide formation process (Powner *et al.* 2009).

We note that besides glyceraldehyde and dihydroxyacetone, other interesting sugars are ribose, ribofuranoside or deoxiribose, for which their rotational spectra have been recently measured and characterized in the laboratory (see Cocinero *et al.* 2012; Peña *et al.* 2013; Ecija *et al.* 2016). Toward astronomical sources with low excitation temperatures of the gas such as the quiescent GMC G+0.693-0.027, the peak of the spectra of these prebiotic COMs shifts to frequencies between 30 and 80 GHz. This frequency range is better suited for the detection of these complex organics because the observed spectra are cleaner from rotational transitions from lighter molecular species, and because the frequency span between transitions is larger, which diminishes the effects of line confusion (see Jimenez-Serra et al. 2014). Therefore, future instrumentation such as the Band 1 and 2 receivers of ALMA, the K and Q bands of the Next Generation Very Large Array (ngVLA), and the Band 5 receivers of the Square Kilometer Array (SKA), will be key in the search of prebiotic COMs in the ISM.

**5. Conclusions**

Since the first prebiotic experiments of Miller-Urey and Joan Oro based on simple molecules detected in the ISM, Astrochemistry has evolved and shown the presence of a wealth of complex organic compounds that can be formed in the ISM with relatively high abundances. The recent discovery in the ISM of prebiotic COMs such as glycolaldehyde, glycolonitrile, cyanomethanimine **and urea**, precursors of ribonucleotides and more complex sugars, might support the hypothesis of a primordial RNA-world in a young Earth. In this paper we present a detailed study of the abundances of the precursors of the proposed RNA-world chemical scheme, toward a massive cloud in the center of our Galaxy (the quiescent GMC G+0.693-0.027) and a star-forming region in the disk of the Milky Way (the hot corino IRAS16293-2422 B). **Our high-sensitivity observations toward the quiescent GMC G+0.693-0.027 has revealed the presence of urea also in this source not affected by star formation with an abundance ~5x10$^{-11}$.** We have also searched for other key molecules in the RNA-world scenario such as 2-amino-oxazole, glyceraldehyde and dihydroxyacetone, toward both astronomical sources, but these searches have yielded no detection so far. We derive the upper limits to the abundance of these species and conclude that, although the ISM is an efficient factory in the production of COMs and of key prebiotic precursors compatible with a primordial RNA-world, the chemical pathways used by the ISM for the formation of these precursors, largely differ those proposed by Powner *et al.* (2009) and Patel *et al.* (2015). In any case, the potentially efficient production in the ISM of these key prebiotic precursors opens up the possibility for the RNA-world prebiotic chemistry to develop elsewhere in the Universe.

**6. Acknowledgements**

We would like to thank Dr. Carlos Briones for illuminating discussions about the primordial RNA-world hypothesis for the origin of life. This work has been supported by MINECO and FEDER (Project number ESP2015-65597-C4-1, CTQ2017-89150-R and ESP2017-86582-C4-1-R), and by the Spanish State Research Agency (AEI) through project number MDM-2017-0737 Unidad de Excelencia "María de Maeztu"- Centro de Astrobiología (INTA-CSIC). We also acknowledge the Basque Government (PIBA 2018-11), the UPV/EHU (PPG17/10) and Fundación BBVA for the financial support. Computational and laser facilities of the UPV/EHU were used in this work. VMR has received funding from the European Union's H2020 research and innovation programme under the Marie Skłodowska-Curie grant agreement No 664931. This paper makes use of the following ALMA data: #2011.0.00007.SV, #2012.1.00712.S, #2013.1.00018.S, #2013.1.00061.S, #2013.1.00278.S, #2013.1.00352.S, and #2015.1.01193.S. ALMA is a partnership of ESO (representing its member states), NSF (USA) and NINS (Japan), together with NRC (Canada), MOST and ASIAA (Taiwan), and KASI (Republic of Korea), in cooperation with the Republic of Chile. The Joint ALMA Observatory is operated by ESO, AUI/NRAO and NAOJ.

**Table Legends**

**Table 1.  Physical parameters derived for the prebiotic molecules from the RNA-world chemical scheme observed toward the hot corino IRAS16293-2422 B.** The lines from all these COMs toward IRAS16293-2422 were identified in the data and synthetic spectra were generated and fitted with the MADCUBA software to constrain the molecular column density, excitation temperature, central radial velocity and linewidth of the emission.

**Table 2.  Physical parameters derived for the prebiotic molecules from the RNA-world chemical scheme observed toward the quiescent GMC G+0.693-0.027.** The same analysis procedure was followed as for Table 1 with the MADCUBA software developed at the Center for Astrobiology in Madrid (CAB).

**Table 3: Spectroscopic constants of the ground vibrational states of 2-amino-oxazole. These constants have been extracted from the experimental work of Møllendal & Konovalov (2010), and they have been used to create the \*.cat files provided in the supplementary material. These files include the information of the rotational transitions of the 0$^+$ and 0$^-$ states of 2-amino-oxazole with their frequencies, energies of the lower level (E$_l$), intensities at 300 K (in logarithmic scale), and quantum numbers.**

**Table 4.  Column density ratios of sugar-like species with respect to formaldehyde (H₂CO).** We have obtained the ratios between the column densities of glycolaldehyde, glycolic acid, glyceraldehyde and dihydroxyacetone with respect to $H_2CO$ for both sources.

**Table 5. Column density ratios of N-bearing COMs with respect to methyl cyanide (HCN).** We have calculated the ratios of the column densities of glycolonitrile, cyanamide, methanimine, urea, cyanomethanimine (E and Z), amino acetonitrile, glycine and 2-amino-oxazole with respect to HCN for both astronomical sources.

**Figure legends**

**Figure 1. Summary of the chemical scheme of the RNA-world scenario.** Extracted from Powner *et al.* (2009), Patel *et al.* (2015), Kitadai *et al.* (2018) and Rivilla *et al.* (2019b; **see the different colors for the arrows). Solid-line boxes indicate molecules that have been detected in space while dotted-line boxes denote those species that remain undetected in space. The goal of this diagram is to visualize at once how many primordial RNA-world prebiotic species have been detected in the ISM, and how they are related to each other.** The names of the molecules shown are: (1) Hydrogen Cyanide; (2) Formaldehyde; (3) Glycolonitrile; (4) Glycolaldehyde; (5) Cyanamide; (6) Glycolic acid; (7) Cyanide; (8) Methanimine; (9) Enol form of glycolaldehyde; (10) Cyanohydrin; (11) Urea; (12) 3-Oxopropanenitrile; (13) Cyanoacetylene; (14) Cyanomethanimine; (15) Aminoacetonitrile; (16) Glyceraldehyde; (17) 2- amino-oxazole; (18) Cytosine; (19) Adenine; (20) Glycine; (21) Dihydroxyacetone (DHA); (22) Glycerol; (23) Beta-ribocytidine-2',3'-cyclic phosphate (pyrimidine ribonucleotide)

**Figure 2. Sample spectra of prebiotic species detected toward the IRAS16293-2422 B hot corino.** Histograms correspond to some of the rotational transitions of COMs detected toward this proto-Sun, while red lines show the Gaussian fits to the observed line profiles performed by the MADCUBA-AUTOFIT tool.

**Figure 3. Sample spectra of prebiotic species detected toward the quiescent GMC G+0.693-0.027.** Histograms correspond to some of the rotational transitions of COMs detected toward this cloud in the Galactic Center, while red lines show the Gaussian fits to the observed line profiles obtained by MADCUBA.

**Figure 4. Clean rotational transitions of urea detected toward G+0.693-0.027.** Our IRAM 30m observations cover 8 clean lines of urea, although three of them are doublets. This translates into five clear urea features detected in the observed spectra (see labeled lines). Red line indicates the spectrum of urea simulated with MADCUBA for a column density of $(6.3\pm0.1) \times 10^{12}$ cm$^{-2}$, $T_{ex}=8$K, $\Delta v=20$ km s$^{-1}$ and $V_{LSR}=69$ km s$^{-1}$.

**Figures.**

**Figure 1. Summary of the chemical scheme of the primordial RNA-world scenario.**

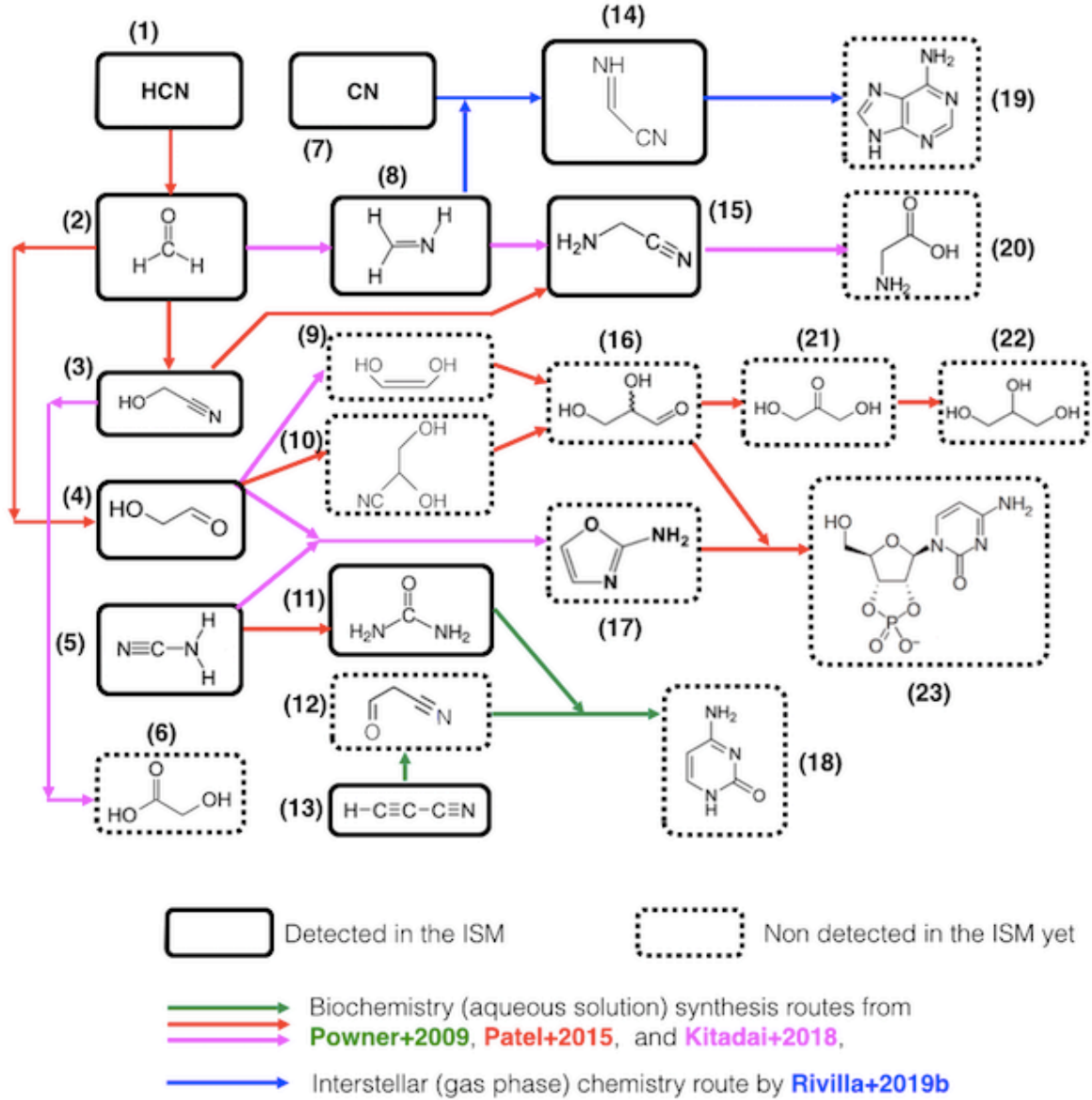

(1) Hydrogen Cyanide; (2) Formaldehyde; (3) Glycolonitrile; (4) Glycolaldehyde; (5) Cyanamide; (6) Glycolic acid; (7) Cyanide; (8) Methanimine; (9) Enol form of glycolaldehyde; (10) Cyanohydrin; (11) Urea; (12) 3-Oxopropanenitrile; (13) Cyanoacetylene; (14) Cyanomethanimine; (15) Aminoacetonitrile; (16) Glyceraldehyde; (17) 2- amino-oxazole; (18) Cytosine; (19) Adenine; (20) Glycine; (21) Dihydroxyacetone (DHA); (22) Glycerol; (23) Beta-ribocytidine-2',3'-cyclic phosphate (pyrimidine ribonucleotide)

**Figure 2. Sample spectra of prebiotic species detected toward the IRAS16293-2422 B hot corino.**

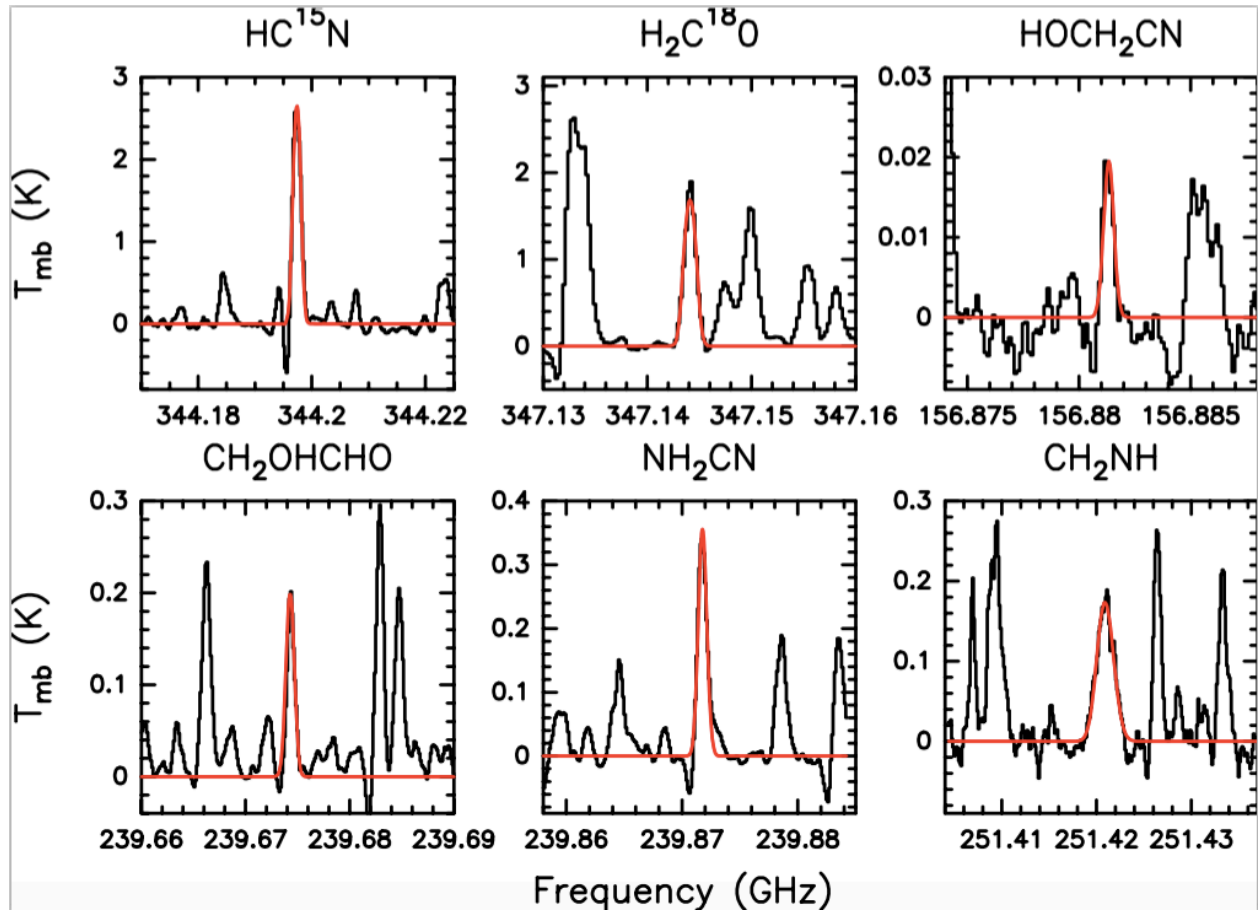

**Figure 3. Sample spectra of prebiotic species detected toward the quiescent GMC G+0.693-0.027.**

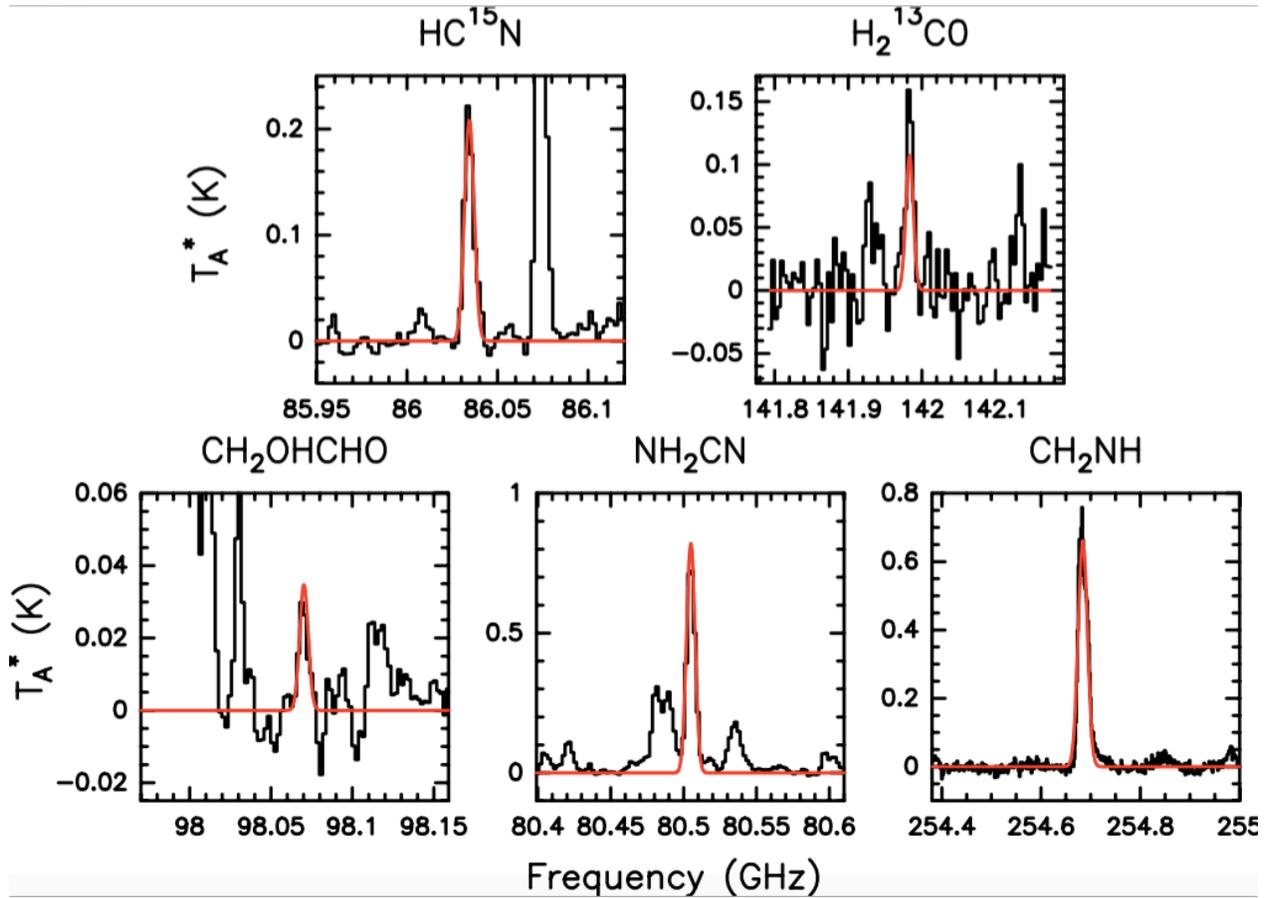

**Figure 4. Clean rotational transitions of urea detected toward G+0.693-0.027.**

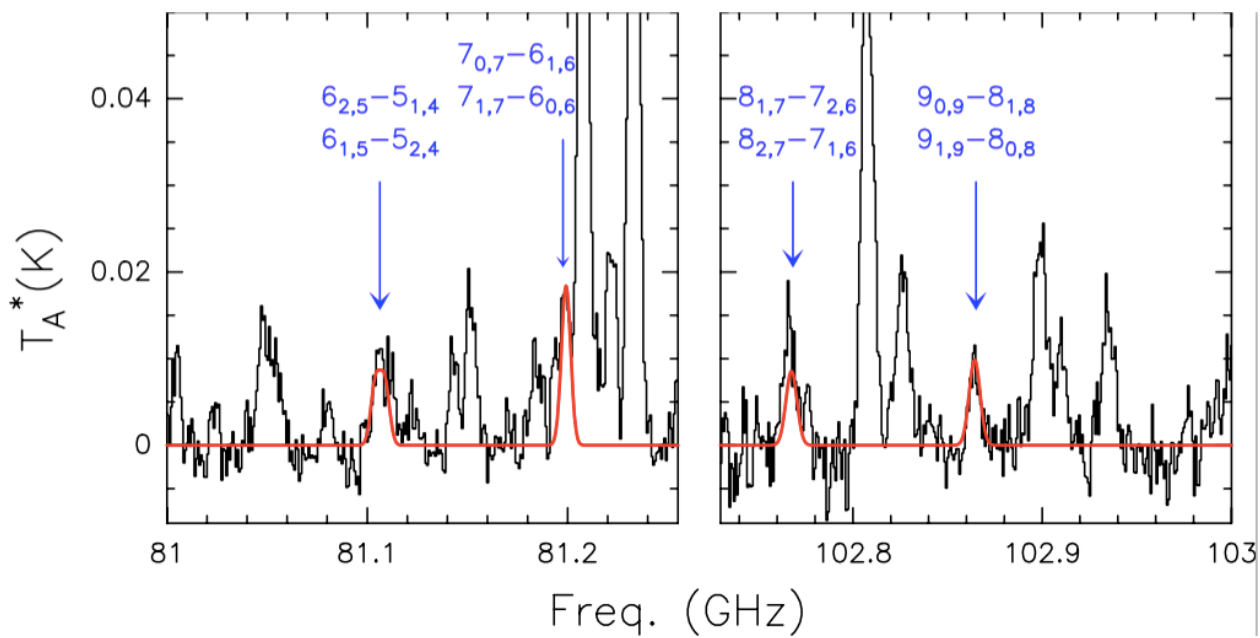

**Table 3: Experimental spectroscopic constants of the ground vibrational states of 2-amino-oxazole used to create *.cat file.**

|  | $0^+$ | $0^-$ |
|---|---|---|
| A/MHz | 9326.4892(16)[1] | 9323.9447(18) |
| B/MHz | 3912.38508(65) | 3907.47653(72) |
| C/MHz | 2760.84451(64) | 2760.31584(65) |
| $\Delta_J$/kHz | 0.29626(50) | 0.29703(52) |
| $\Delta_{JK}$/kHz | 1.1527(55) | 1.1568(59) |
| $\Delta_K$/kHz | 1.7972(26) | 1.7881(27) |
| $\delta_J$/kHz | 0.08323(45) | 0.08220(41) |
| $\delta_K$/kHz | 1.1305(97) | 1.141(10) |
| $N_{lines}$ | 527 | 458 |

[1]The numbers in parentheses are 1σ uncertainties in units of the last decimal digit.

**Table 1. Physical parameters derived for the prebiotic molecules from the RNA-world chemical scheme observed toward the hot corino IRAS16293-2422 B.**

| Molecule Name | Formula | $T_{ex}$ (K) | $V_{LSR}$ (km s$^{-1}$) | $\Delta v$ (km s$^{-1}$) | N(cm$^{-2}$) | $\chi^d$ | Reference |
|---|---|---|---|---|---|---|---|
| Hydrogen cyanide ($^{15}$N) | HC$^{15}$N | 150[a] | 2.42±0.03 | 1.29±0.05 | ≥3.5x10$^{14e}$ | ≥1.3x10$^{-11}$ | This work; see also Takakuwa *et al.* (2007) |
| Formaldehyde ($^{18}$O) | H$_2$C$^{18}$O | 99±11 | 2.65±0.02 | 0.98±0.06 | (3.5±0.3)x10$^{15}$ | 1.3x10$^{-10}$ | This work; see also Persson *et al.* (2018) |
| Glycolonitrile (hot) | HOCH$_2$CN | 158±38 | 2.62±0.15 | 1.03±0.03[b] | (1.8±0.1)x10$^{15b}$ | 6.5x10$^{-11}$ | Zeng *et al.* (2019) |
| Glycolaldehyde | CH$_2$(OH)CHO | 284-300 | … | … | (3.3-6.8)x10$^{16}$ | (1.2-2.4)x10$^{-9}$ | From Jorgensen *et al.* (2016); Rivilla *et al.* 2019a |
| Cyanamide (hot) | NH$_2$CN | 290±40 | 2.76±0.04 | 1.03[b,c] | (5.2±0.6)x10$^{14b}$ | 1.9x10$^{-11}$ | This work; see also Coutens *et al.* (2018) |
| Methanimine | CH$_2$NH | 150[a] | 3.3±0.3 | 2.0±0.9 | (4.0±0.8)x10$^{15}$ | 1.4x10$^{-10}$ | This work; see also Ligterink *et al.* (2018) |
| Glycolic Acid | HOCOCH$_2$OH | 150[a] | … | … | ≤1.2x10$^{14}$ | ≤4.3x10$^{-12}$ | This work |
| Urea | NH$_2$CONH$_2$ | 150[a] | … | … | ≤6.5x10$^{14}$ | ≤2.3x10$^{-11}$ | This work |
| E-Cyanomethanimine | E-HNCHCN | 150[a] | … | … | ≤8.7x10$^{13}$ | ≤3.1x10$^{-12}$ | This work |
| Z-Cyanomethanimine | Z-HNCHCN | 150[a] | … | … | ≤1.0x10$^{15}$ | ≤3.6x10$^{-11}$ | This work |
| Amino acetonitrile | NH$_2$CH$_2$CN | 150[a] | … | … | ≤1.2x10$^{14}$ | ≤4.3x10$^{-12}$ | This work |
| Glyceraldehyde | C$_3$H$_6$O$_3$ | 150[a] | … | … | ≤1.6x10$^{15}$ | ≤5.7x10$^{-11}$ | This work |
| Dihydroxyacetone (DHA) | C$_3$H$_6$O$_3$ | 150[a] | … | … | ≤1.1x10$^{16}$ | ≤3.9x10$^{-10}$ | This work |
| Glycine (Conformer I) | NH$_2$CH$_2$COOH | 150[a] | … | … | ≤5.8x10$^{15}$ | ≤2.1x10$^{-10}$ | This work |
| 2-amino-oxazole 0$^+$ | C$_3$H$_4$N$_2$O 0$_+$ | 150[a] | … | … | ≤3.2x10$^{14}$ | ≤1.1x10$^{-11}$ | This work |

| 2-amino-oxazole 0$^-$ | $C_3H_4N_2O$ 0$^-$ | 150[a] | … | … | ≤1.1x10$^{15}$ | ≤3.9x10$^{-11}$ | This work |

[a], Tex fixed to this value within MADCUBA-AUTOFIT analysis tool; [b], Molecule required two temperature components to fit the observed spectra; [c], Linewidth had to be fixed for the MADCUBA-AUTOFIT tool to converge; [d], Molecular abundances with respect to $H_2$ calculated assuming $N(H_2)$=2.8x10$^{25}$ cm$^{-2}$ (Martin-Domenech *et al.* 2017); e, The HC$^{15}$N line is optically thick (optical depth of 1.4) and thus its derived column density is a lower limit.

**Table 2. Physical parameters derived for the prebiotic molecules from the RNA-world chemical scheme observed toward the quiescent GMC G+0.693-0.027.**

| Molecule Name | Formula | $T_{ex}$ (K) | $V_{LSR}$ (km s$^{-1}$) | $\Delta v$ (km s$^{-1}$) | N (cm$^{-2}$) | $\chi^f$ | Reference |
|---|---|---|---|---|---|---|---|
| Hydrogen cyanide ([15]N) | HC[15]N | 4.7±0.3 | 71.6±0.5 | 22±1 | (1.1±0.1)x10$^{13}$ | 3.0x10$^{-10}$ | From Zeng *et al.* (2018) |
| Formaldehyde ([13]C) | H$_2^{13}$CO | 6.6±0.1 | 70.4±0.1 | 19.8±0.2 | (2.00±0.03)x10$^{13}$ | 1.5x10$^{-10}$ | This paper |
| Glycolonitrile | HOCH$_2$CN | 8[c] | … | 20[d] | ≤7.7x10$^{12}$ | ≤5.7x10$^{-11}$ | This paper |
| Glycolaldehyde | CH$_2$(OH)CHO | 8[c] | 70[d] | 20[d] | (3.31±0.04)x10$^{13}$ | 2.5x10$^{-10}$ | This paper |
| Cyanamide | NH$_2$CN | 6.6[a] | 67[a] | 24[a] | (3.1±0.2)x10$^{14b}$ | 2.3x10$^{-9}$ | From Zeng *et al.* (2018) |
| Methanimine | CH$_2$NH | 9.7±0.4 | 69±1 | 25±1 | (5.4±0.3)x10$^{14}$ | 4.3x10$^{-9}$ | From Zeng *et al.* (2018) |
| Glycolic Acid | HOCOCH$_2$OH | 8[c] | … | 20[d] | ≤2.3x10$^{13}$ | ≤1.7x10$^{-10}$ | This paper |
| Urea | NH$_2$CONH$_2$ | 8[c] | … | 20[d] | (6.3±0.1)x10$^{12}$ | 4.7x10$^{-11}$ | This paper |
| E-Cyanomethanimine | E-HNCHCN | 8[c] | 68.0±0.8 | 21±2 | (0.33±0.03)x10$^{14}$ | 2.4x10$^{-10}$ | From Rivilla *et al.* (2019) |
| Z-Cyanomethanimine | Z-HNCHCN | 8±2 | 68.3±0.8 | 20[d] | (2.0±0.6)x10$^{14}$ | 1.5x10$^{-9}$ | From Rivilla *et al.* (2019) |
| Amino acetonitrile | NH$_2$CH$_2$CN | 15[e] | … | 20[d] | ≤0.6x10$^{13}$ | ≤4.7x10$^{-11}$ | From Zeng *et al.* (2018) |
| Glyceraldehyde | C$_3$H$_6$O$_3$ | 8[c] | … | 20[d] | ≤1.3x10$^{13}$ | ≤9.3x10$^{-11}$ | This paper |
| Dihydroxyacetone (DHA) | C$_3$H$_6$O$_3$ | 8[c] | … | 20[d] | ≤6.8x10$^{12}$ | ≤5.0x10$^{-11}$ | This paper |
| Glycine (Conformer I) | NH$_2$CH$_2$COOH | 8[c] | … | 20[d] | ≤5.6x10$^{13}$ | ≤4.1x10$^{-10}$ | This paper |
| 2-amino-oxazole 0$^+$ | C$_3$H$_4$N$_2$O 0$^+$ | 8[c] | … | 20[d] | ≤1.1x10$^{13}$ | ≤8.1x10$^{-11}$ | This paper |

| 2-amino-oxazole 0[-] | $C_3H_4N_2O$ 0[-] | $8$[c] | … | $20$[d] | $\leq 1.1 \times 10^{13}$ | $\leq 8.1 \times 10^{-11}$ | This paper |

[a], Average values between the ortho and para forms of $NH_2CN$ (see also Zeng *et al.* 2018); [b], Total column density calculated adding the column densities derived from the ortho and para forms of $NH_2CN$; [c], Temperature fixed to the Tex inferred for Z-cyanomethanimine, so that the MADCUBA-AUTOFIT tool could converge; [d], Linewidth and $V_{LSR}$ fixed to the typical value of the molecular emission in G+0.693-0.027 so that MADCUBA-AUTOFIT could coverge; [e], As assumed in Zeng *et al.* (2018); [f], Molecular abundances with respect to $H_2$ calculated assuming $N(H_2)=1.35 \times 10^{23}$ $cm^{-2}$ (Martin *et al.* 2008).

**Table 4. Column density ratios of sugar-like species with respect to formaldehyde ($H_2CO$).**

| Molecule/Formaldehyde[a] | Glycolaldehyde | Glycolic Acid | Glyceraldehyde | Dihydroxyacetone |
|---|---|---|---|---|
| IRAS16293-2422 | 0.01-0.02 | $\leq 4 \times 10^{-5}$ | $\leq 6 \times 10^{-4}$ | $\leq 0.004$ |
| G+0.693-0.027 | 0.08 | $\leq 0.06$ | $\leq 0.03$ | $\leq 0.017$ |

[a], The column density of $H_2CO$ toward IRAS16293 is $2.8 \times 10^{18}$ $cm^{-2}$ and toward G+0.693-0.027, it is $4 \times 10^{14}$ $cm^{-2}$ (see text in Section 3.2 for the details on how these column densities have been calculated).

**Table 5. Column density ratios of N-bearing COMs with respect to methyl cyanide (HCN).**

| Molecule/HCN[a] | Glycolonitrile | Cyanamide | Methanimine | Urea | Amino Acetonitrile | Cyanomethanimine | Glycine | 2-amino-oxazole |
|---|---|---|---|---|---|---|---|---|
| IRAS16293-2422 | ~0.03 | ~0.009 | ~0.07 | $\leq 0.01$ | $\leq 0.002$ | $\leq 0.002$-0.02 | $\leq 0.1$ | $\leq 0.005$-0.02 |
| G+0.693-0.027 | $\leq 0.001$ | 0.05 | 0.08 | 0.001 | $\leq 0.0009$ | 0.005-0.03 | $\leq 0.009$ | $\leq 0.002$ |

[a], Column densities extracted from Tables 1 and 2; the estimated column densities for HCN are ~$6 \times 10^{16}$ $cm^{-2}$ toward IRAS16293-2422 B and >$6.6 \times 10^{15}$ $cm^{-2}$ toward G+0.693-0.027.

**Table 6. Molecular rotational transitions observed toward the hot corino IRAS16293-2422 B and reported in this work.**

| Molecule Name | Formula | Transition | Freq (MHz) | $Log_{10}[I(300K)]$ | $E_l$ $(cm^{-1})$ | $\Delta v$ $(km\ s^{-1})$ | $v_{LSR}(km\ s^{-1})$ | Area $(Jy\ kms^{-1})$ |
|---|---|---|---|---|---|---|---|---|
| Hydrogen cyanide ($^{15}$N) | $HC^{15}N$ | 4-3 | 344200.109 | -2.72583 | 24.8 | 1.29±0.05 | 2.42±0.03 | 4.45±0.09 |
| Formaldehyde ($^{18}$O) | $H_2C^{18}O$ | $5_{0,5}-4_{0,4}$ | 345881.039 | -2.2054 | 23.1 | 0.98±0.06[b] | 2.60±0.01[b] | 2.98±0.06 |
| | | $5_{3,2}-4_{3,1}$ | 347144.011 | -2.0733 | 97.3 | 0.98±0.06[b] | 2.60±0.01[b] | 2.55±0.09 |
| | | $5_{2,3}-4_{2,2}$ | 348032.433 | -2.3444 | 56.2 | 0.98±0.06[b] | 2.60±0.01[b] | 2.05±0.03 |
| | | $5_{2,4}-4_{2,3}$ | 346869.178 | -2.3472 | 56.1 | 0.98±0.06[b] | 2.60±0.01[b] | 2.03±0.03 |
| | | $5_{4,1}-4_{4,0}$ | 346984.094 | -2.9202 | 155.0 | 0.98±0.06[b] | 2.60±0.01[b] | 0.26±0.02 |
| | | $5_{4,2}-4_{4,1}$ | 346984.067 | -2.9202 | 155.0 | 0.98±0.06[b] | 2.60±0.01[b] | 0.26±0.02 |
| Cyanamide (hot) | $NH_2CN$ | $12_{0,12}-11_{0,11}$ | 239682.9192 | -2.0556 | 93.5 | 1.03[a] | 2.75±0.04[b] | 0.20±0.01 |
| | | $12_{2,10}-11_{2,9}$ | 239431.6816 | -2.1594 | 132.8 | 1.03[a] | 2.75±0.04[b] | 0.16±0.01 |
| | | $12_{0,12}-11_{0,11}$ | 239872.0312 | -2.4181 | 44.0 | 1.03[a] | 2.75±0.04[b] | 0.09±0.01 |
| | | $12_{4,9}-11_{4,8}$ | 239562.2461 | -2.4349 | 250.4 | 1.03[a] | 2.75±0.04[b] | 0.08±0.01 |
| | | $12_{4,8}-11_{4,7}$ | 239562.2462 | -2.4349 | 250.4 | 1.03[a] | 2.75±0.04[b] | 0.08±0.01 |
| | | $12_{2,11}-11_{2,10}$ | 239858.9859 | -2.5145 | 84.3 | 1.03[a] | 2.75±0.04[b] | 0.07±0.01 |
| | | $12_{5,7}-11_{5,6}$ | 239702.4999 | -2.5467 | 295.0 | 1.03[a] | 2.75±0.04[b] | 0.06±0.01 |
| | | $12_{5,8}-11_{5,7}$ | 239702.4999 | -2.5467 | 295.0 | 1.03[a] | 2.75±0.04[b] | 0.06±0.01 |
| | | $12_{3,9}-11_{3,8}$ | 239613.4863 | -2.7462 | 181.9 | 1.03[a] | 2.75±0.04[b] | 0.04±0.01 |

| Molecule | Formula | Transition | Frequency | | | | | |
|---|---|---|---|---|---|---|---|---|
| | | $12_{3,10}-11_{3,9}$ | 239613.4863 | -2.7462 | 181.9 | $1.03^a$ | $2.75\pm0.04^b$ | $0.04\pm0.01$ |
| | | $12_{4,9}-11_{4,8}$ | 239805.2745 | -2.8038 | 204.9 | $1.03^a$ | $2.75\pm0.04^b$ | $0.04\pm0.01$ |
| | | $12_{4,8}-11_{4,7}$ | 239805.2745 | -2.8038 | 204.9 | $1.03^a$ | $2.75\pm0.04^b$ | $0.04\pm0.01$ |
| Methanimine | $CH_2NH$ | $6_{0,6}-5_{1,5}$ | 251421.178 | -2.8996 | 36.1 | $2.0\pm0.9$ | $3.3\pm0.3$ | $0.44\pm0.04$ |
| Glycolic Acid | $HOCOCH_2OH$ | $37_{4,33}-36_{4,32}$ | 247406.9484 | -3.5061 | 161.0 | $1.03^a$ | $2.52^a$ | $\leq0.003$ |
| Urea | $NH_2CONH_2$ | $14_{1,14}-13_{0,13}$ | 157024.1234 | -3.1113 | 35.2 | $1.03^a$ | $2.52^a$ | $\leq0.03$ |
| | | $14_{0,14}-13_{1,13}$ | 157024.1234 | -3.1113 | 35.2 | $1.03^a$ | $2.52^a$ | $\leq0.03$ |
| E-Cyanomethanimine | E-HNCHCN | $25_{4,22}-24_{4,21}$ | 239583.436 | -2.6463 | 126.7 | $1.03^a$ | $2.52^a$ | $\leq0.013$ |
| Z-Cyanomethanimine | Z-HNCHCN | $25_{4,21}-24_{4,20}$ | 243136.2686 | -3.4159 | 123.5 | $1.03^a$ | $2.52^a$ | $\leq0.03$ |
| Amino Acetonitrile | $NH_2CH_2CN$ | $38_{3,36}-37_{3,35}$ | 342845.8958 | -2.7264 | 220.4 | $1.03^a$ | $2.52^a$ | $\leq0.013$ |
| Glyceraldehyde | $C_3H_6O_3$ | $24_{7,17}-23_{6,17}$ | 157134.1516 | -4.3181 | 51.9 | $1.03^a$ | $2.52^a$ | $\leq0.005$ |
| Dihydroxyacetone | $C_3H_6O_3$ | $26_{7,20}-26_{6,21}$ | 101268.1871 | -5.23 | 54.0 | $1.03^a$ | $2.52^a$ | $\leq0.003$ |
| Glycine | $NH_2CH_2COOH$ | $35_{9,26}-34_{9,25}$ | 252275.3207 | -4.1826 | 156.7 | $1.03^a$ | $2.52^a$ | $\leq0.02$ |
| 2-amino-oxazole $0^+$ | $C_3H_4N_2O$ $0^+$ | $19_{19,1}-18_{18,0}$ | 348360.1687 | -3.7665 | 102.8 | $1.03^a$ | $2.52^a$ | $\leq0.006$ |
| | | $19_{19,0}-18_{18,1}$ | 348360.1687 | -3.7665 | 102.8 | $1.03^a$ | $2.52^a$ | $\leq0.006$ |
| 2-amino-oxazole $0^-$ | $C_3H_4N_2O$ $0^-$ | $28_{14,14}-27_{13,15}$ | 348375.6632 | -3.9458 | 118.8 | $1.03^a$ | $2.52^a$ | $\leq0.012$ |
| | | $28_{14,15}-27_{13,14}$ | 348375.4669 | -3.9458 | 118.8 | $1.03^a$ | $2.52^a$ | $\leq0.012$ |

**Table 7. Molecular rotational transitions observed toward the quiescent GMC G+0.693-0.027 and reported in this work.**

| Molecule Name | Formula | Transition | Freq (MHz) | $Log_{10}[I(300K)]$ | $E_l$ (cm⁻¹) | $\Delta v$ (km s⁻¹) | $v_{LSR}$(km s⁻¹) | Area (K kms⁻¹) |
|---|---|---|---|---|---|---|---|---|
| Formaldehyde ($^{13}$C) | $H_2^{13}CO$ | $2_{0,2}$-$1_{0,1}$ | 141983.7404 | -3.3163 | 2.4 | 19.8±0.2[b] | 70.4±0.1[b] | 4.16±0.09 |
| | | $3_{0,3}$-$2_{0,2}$ | 212811.184 | -2.801 | 7.1 | 19.8±0.2[b] | 70.4±0.1[b] | 2.21±0.04 |
| Glycolonitrile | $HOCH_2CN$ | $8_{1,7}$-$7_{1,6}$ | 75463.333 | -4.632 | 9.8 | … | … | ≤0.05 |
| Glycolaldehyde | $CH_2(OH)CHO$ | $3_{3,0}$-$2_{2,1}$ | 98070.5113 | -4.6328 | 2.8 | 20[a] | 70[a] | 0.45±0.2 |
| | | $3_{3,1}$-$2_{2,0}$ | 97919.5791 | -4.6335 | 2.9 | 20[a] | 70[a] | 0.45±0.2 |
| | | $4_{3,1}$-$3_{2,2}$ | 109877.1408 | -4.5319 | 4.0 | 20[a] | 70[a] | 0.40±0.2 |
| | | $5_{4,1}$-$4_{3,2}$ | 146445.05 | -4.1447 | 7.7 | 20[a] | 70[a] | 0.32±0.2 |
| | | $5_{2,3}$-$4_{1,4}$ | 107886.2448 | -4.9176 | 4.0 | 20[a] | 70[a] | 0.32±0.2 |
| Glycolic Acid | $HOCOCH_2OH$ | $11_{2,10}$-$10_{2,9}$ | 74028.9895 | -4.7706 | 19.0 | 20[a] | 70[a] | ≤0.05 |
| Urea | $NH_2CONH_2$ | $7_{1,7}$-$6_{0,6}$ | 81199.2 | -3.9453 | 8.7 | 20[a] | 69[a] | 0.20±0.02 |
| | | $7_{0,7}$-$6_{1,6}$ | 81199.2 | -3.9453 | 8.7 | 20[a] | 69[a] | 0.20±0.02 |
| | | $6_{1,5}$-$5_{2,4}$ | 81104.13 | -4.1074 | 7.9 | 20[a] | 69[a] | 0.15±0.02 |
| | | $6_{2,5}$-$5_{1,4}$ | 81108.77 | -4.1074 | 7.9 | 20[a] | 69[a] | 0.15±0.02 |
| | | $9_{1,9}$-$8_{0,8}$ | 102864.32 | -3.6353 | 14.4 | 20[a] | 69[a] | 0.11±0.02 |
| | | $9_{0,9}$-$8_{1,8}$ | 102864.32 | -3.6353 | 14.4 | 20[a] | 69[a] | 0.11±0.02 |

| | | | | | | | |
|---|---|---|---|---|---|---|---|
| | | $8_{2,7}-7_{1,6}$ | 102767.56 | -3.7539 | 13.7 | 20[a] | 69[a] | 0.09±0.02 |
| | | $8_{1,7}-7_{2,6}$ | 102767.56 | -3.7539 | 13.7 | 20[a] | 69[a] | 0.09±0.02 |
| Glyceraldehyde | $C_3H_6O_3$ | $8_{7,1}-7_{6,1}$ | 78879.586 | -5.1099 | 8.3 | 20[a] | 69[a] | ≤0.02 |
| | | $8_{7,2}-7_{6,2}$ | 78879.586 | -5.1099 | 8.3 | 20[a] | 69[a] | ≤0.02 |
| Dihydroxyacetone (DHA) | $C_3H_6O_3$ | $5_{5,1}-4_{4,0}$ | 90104.154 | -5.714 | 5.5 | 20[a] | 69[a] | ≤0.014 |
| | | $5_{5,0}-4_{4,1}$ | 90104.154 | -5.714 | 5.5 | 20[a] | 69[a] | ≤0.014 |
| Glycine (Conformer I) | $NH_2CH_2COOH$ | $4_{4,1}-3_{3,0}$ | 75795.2729 | -6.2317 | 3.4 | 20[a] | 69[a] | ≤0.03 |
| 2-amino-oxazole $0^+$ | $C_3H_4N_2O$ $0^+$ | $5_{4,2}-4_{3,1}$ | 75299.4615 | -5.6144 | 4.0 | 20[a] | 69[a] | ≤0.03 |
| 2-amino-oxazole $0^-$ | $C_3H_4N_2O$ $0^-$ | $5_{4,2}-4_{3,1}$ | 75273.6888 | -5.6146 | 4.0 | 20[a] | 69[a] | ≤0.03 |

[a], Linewidth and $V_{LSR}$ fixed in MADCUBA; [b], Linewidths and $V_{LSR}$ are the same for all transitions from the same molecule when simulating the spectra within MADCUBA